\documentclass{basi}

\usepackage{graphicx}

\begin{document}
\title[SZ Science with an ALMA Band 1 Receiver System]{SZ Science with an ALMA Band 1 Receiver System} 
\author[Anna M. M. Scaife and Keith J. B. Grainge]
       {Anna M. M. Scaife$^1$\thanks{e-mail:ascaife@cp.dias.ie} and
	Keith J. B. Grainge$^{2,3}$ \\ 
        $^1$ Dublin Institute for Advanced Studies, 31 Fitzwilliam Place, Dublin 2, Ireland\\
	$^2$ Astrophysics Group, Cavendish Laboratory, J. J. Thomson Avenue, Cambridge, CB3 0HE, UK\\
	$^3$ Kavli Institute for Cosmology, Madingley Road, Cambridge, UK\\}
\date{Received 2010 February 8; accepted 2010 November 24}
\maketitle

\label{firstpage}
\begin{abstract}
We present the first full interferometric simulations of galaxy clusters containing radio plasma bubbles as observed by the proposed Q band receiver system for the ALMA telescope. We discuss the observational requirements for detecting intracluster substructure directly from the SZ signal and the advantages of these observations over those made with the current generation of SZ survey instruments.
\end{abstract}

\section{Introduction}
\label{sec:intro}
Observations of the Sunyaev--Zel'dovich (SZ; see e.g. Birkinshaw 1999) effect have become common over the past decade and 
the cosmological advances which can be gained from surveys of galaxy clusters through this effect are now being 
exploited by the current generation of SZ survey telescopes such as the Arcminute Microkelvin Imager (AMI Consortium: 
Zwart et~al. 2008), Sunyaev-Zel'dovich Array (SZA; Carlstrom 2006), South Pole Telescope (SPT; Ruhl et~al. 2004) and 
the Atacama Cosmology Telescope (ACT; Fowler 2004). In addition to being an almost redshift independent effect, 
the SZ effect provides a direct measure of the integrated line of sight pressure through a cluster. The next stage 
in SZ science is to exploit this fact in order to investigate the substructure of intra-cluster gas. 

Recent X-ray observations with the XMM (Jansen et~al. 2001) and Chandra (Weisskopf 1999) satellites have revealed the presence of cavities in the gas within clusters of galaxies. Notable amongst these clusters is the Perseus cluster (B{\"o}hringer et~al. 1993; Fabian et~al. 2000), which provides an excellent example of a nearby ($z=0.0178$) cluster with large cavities close to the centre of the electron gas density distribution. It has been proposed that these cavities within the gas of galaxy clusters are bubbles containing the relic plasma from the jets of radio galaxies. Understanding the composition of this plasma is a key question in astrophysics. However this hot relativistic gas is difficult to detect through X-ray observations due to its low density, since X-ray emissivity is proportional to the square of the gas density. Since the SZ effect is linearly proportional to the density this makes it a far better probe.

Current SZ instruments are designed to observe clusters in their entirety, in order to probe the mass and number evolution of clusters for cosmological parameter estimation (e.g. Kneissl et~al. 2001). In order to investigate the substructure within clusters telescopes with higher resolution and surface brightness sensitivity are required. A Q band instrument on the ALMA telescope will provide these requirements. In this paper we simulate observations of a Perseus-like galaxy cluster for the proposed instrument in order to evaluate and demonstrate the viability of such an experiment. In Section~\ref{sec:clmod} of this paper we introduce the SZ effect and describe our simulated cluster, and in Section~\ref{sec:intsim} we present the results of our interferometric simulations. In Section~\ref{sec:disc} we discuss the implications of these results for ALMA and compare them with similar observations from current SZ telescopes.

\section{Cluster Modelling}
\label{sec:clmod}

The Sunyaev-Zel'dovich (SZ) effect is caused by the inverse Compton scattering of cosmic microwave background (CMB) photons as 
they pass through the hot electron gas (typically $T_{\rm{e}} = 10^8$\,K) in clusters of galaxies. The change in intensity of the CMB signal due to the thermal gas is generally parameterized as
\begin{equation}
\Delta I(x) = g(x)y_{\rm{gas}} 
\end{equation} 
where,
\begin{equation}
g(x) = \frac{x^4 \exp{x}}{(\exp{x}-1)^2}\left(x\frac{\exp{x}+1}{\exp{x}-1} - 4\right).
\end{equation}
$x$ is the dimensionless quantity $h\nu/kT_0$, where $h$ is the Planck constant, $\nu$ is the frequency of observation, $k$ is the Boltzmann constant and $T_0=2.726$\,K is the temperature of the CMB. Additional changes are also caused by the proper motion of the cluster (kinetic SZ effect) and the presence of relativistic gas. These changes are an order of magnitude and several orders of magnitude smaller than the thermal effect, respectively. The quantity $y_{\rm{gas}}$, known as the Componization y-parameter, is defined as the integrated pressure along the line of sight,
\begin{equation}
y_{\rm{gas}} = \frac{\sigma_T}{m_{\rm{e}}c^2}\int{n_{\rm{e}} k T_{\rm{e}} d\ell }.
\end{equation}
For the purposes of this study the density profile of relaxed clusters in hydrostatic equilibrium is generally considered to be well described by an isothermal $\beta$-model (Cavaliere \& Fusco-Femiano 1976; 1978) and we may parameterize the electron gas density profile as
\begin{equation} 
n_{\rm{e}}(r) = n_0\left(1+\frac{r^2}{r_c^2}\right)^{-3\beta},
\end{equation}
where $r$ is the distance from the centre of the cluster, $r_c$ is a characteristic core radius, $n_0$ is the central electron density and $\beta$ is an index which varies typically between 0.67 and 1.5.
The corresponding y-parameter at projected radius $\rho$ is consequently,
\begin{equation}
y(\rho) = y_0 B(0.5,\frac{3\beta}{2}-0.5) \left(1+\frac{\rho^2}{r_c^2}\right)^{-(3\beta-1)/2},
\end{equation}
where $y_0$ is the central y-parameter, and $B$ is the incomplete Beta function. Often multiple $\beta$ models will be employed to describe an observed cluster profile. A cluster profile fited with N $\beta$ models will be expressed as 
\begin{equation}
y_{\rm{tot}}(\rho) = \sum_{i=1}^{\rm{N}}{y_{0,i}B(0.5,\frac{3\beta_i}{2}-0.5)\left(1+\frac{\rho^2}{r_{c,i}^2}\right)^{-(3\beta_i-1)/2}},
\end{equation}
where each of the component $\beta$ models possess their own central y-parameter, $y_i$, core radius $r_{c,i}$ and $\beta$ index, $\beta_i$. 

\begin{figure}
\centerline{\includegraphics[width=0.8\textwidth]{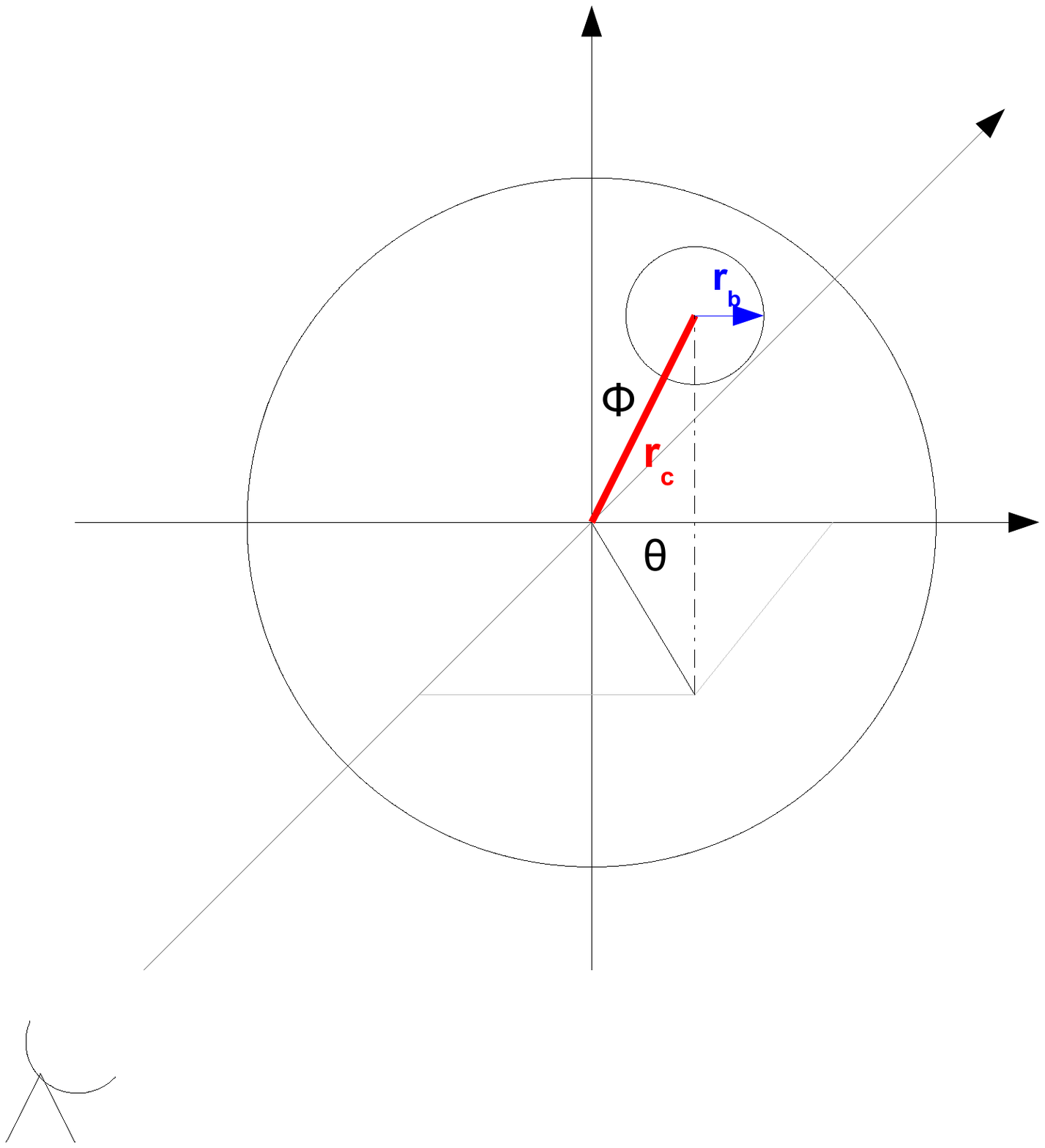}}
\caption{Geometry of a bubble or cavity within a galaxy cluster. The labels $r_c$ and $r_b$ denote the distance of the centre of the bubble from the centre of the cluster and the radius of the bubble itself, respectively. The angles $\theta$ and $\phi$ indicate the inclination angles of the centre of the bubble from the plane of the sky, which is perpendicular to the line of sight of the telescope. \label{fig:bubgeom}}
\end{figure}

\begin{figure}
\centerline{\includegraphics[width=0.5\textwidth]{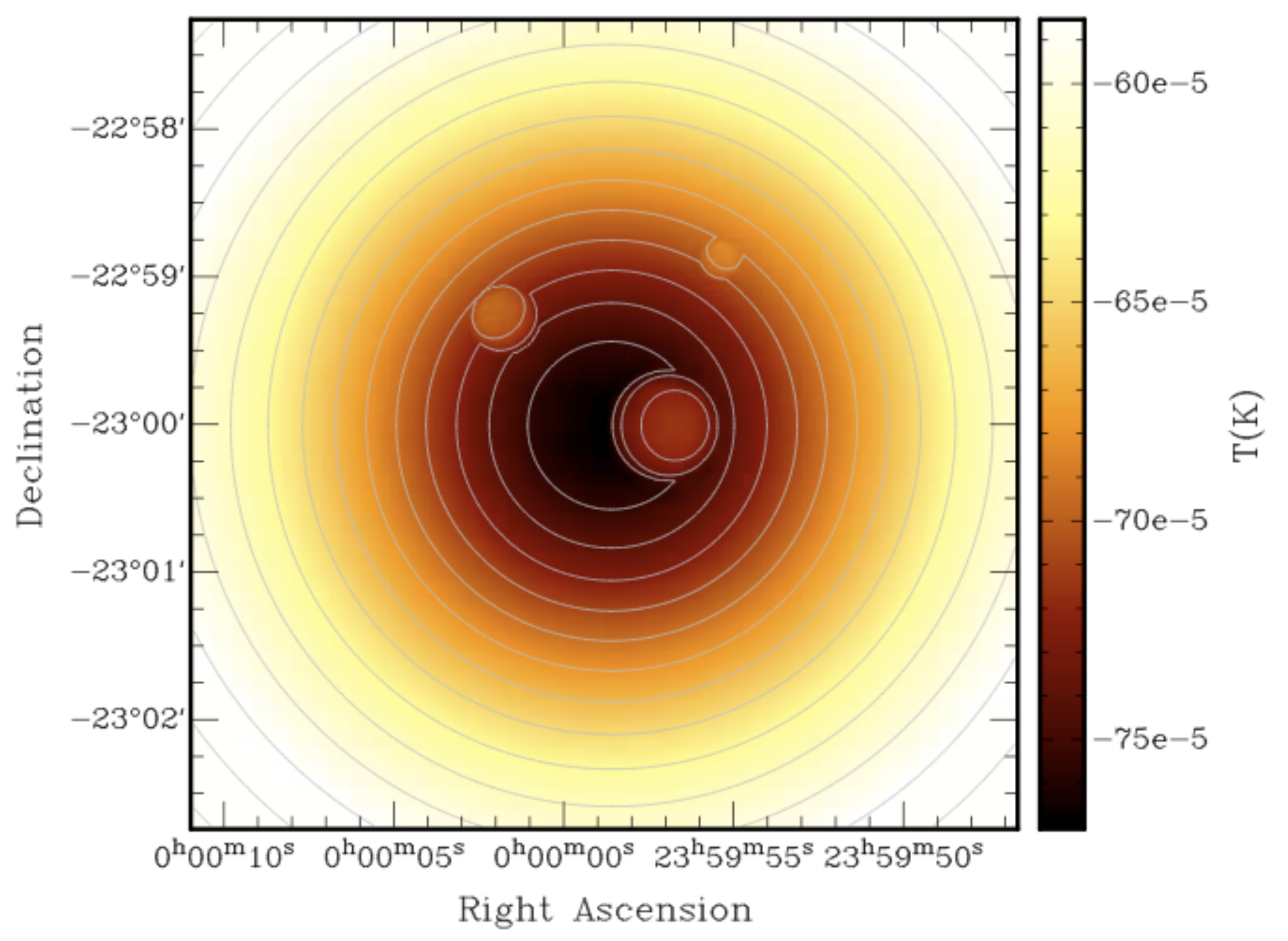}\qquad \includegraphics[width=0.5\textwidth]{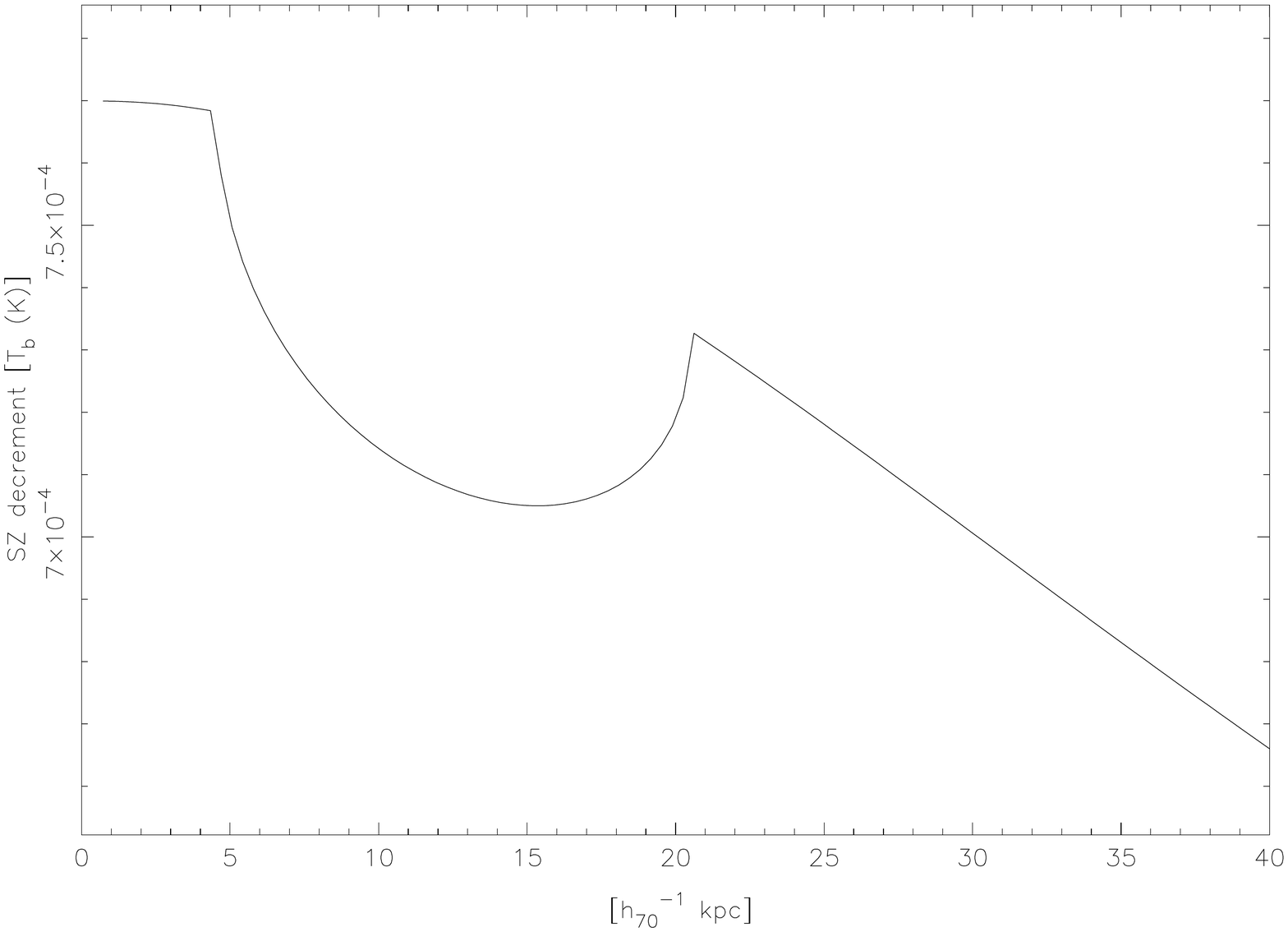}}

\centerline{(a) \hspace{0.5\textwidth} (b)}
\caption{Simulated cluster containing 3 bubbles. The positions and sizes of the bubbles are listed in Table~\ref{tab:bubbles}.\label{fig:bubbles} 
Figure (a) shows a relief of the brightness temperature of the simulated cluster and (b) shows a cut taken through the largest of the three bubbles.}
\end{figure}

It has been observed in X-ray data that clusters of galaxies often possess distinct substructure in the hot electron gas. 
This substructure often appears to be bubble-like in morphology. If we assume spherical bubbles within our cluster we can 
parameterize them as shown in Fig.~\ref{fig:bubgeom}. We consider that the relativistic gas within these bubbles contributes 
negligibly to the SZ emission of the cluster. Details of the contributions from different plasma compositions to the integrated 
SZ effect are described in Pfrommer, En{\ss}lin \& Sarazin (2005) and this assumption is consistent with their Scenario 1, 
which is most widely used. Since any contribution to the SZ effect from the bubble plasma will reduce the contrast to the overall 
cluster, this may be considered the optimal case for detection. Under these circumstances an analytic integral for the line 
of sight y-parameter through a cluster containing these types of bubbles, $y_{\rm{cl}}$, was derived by Pfrommer et~al. (2005) to be
\begin{eqnarray}
\nonumber y_{\rm{cl}}(x_1,x_2) &=& y_{\rm{tot}}(x_1,x_2)\\
&-& \sum_{i=1}^{N}{y_i\left(1+\frac{x_1^2+x_2^2}{r_{c,i}^2}\right)^{-(3\beta_{i}-1)/2}\times\left[\frac{\rm{sgn}(z)}{2} I_{q_{i}(z)}(0.5,1.5\beta_i-0.5)\right]_{z_{-}}^{z_{+}}}.
\end{eqnarray}
Where a line of sight at position $(x_1,x_2)$ will intersect the bubble at $z_{\pm}$, where
\begin{equation}
z_{\pm} = r_c\sqrt{1-\cos \theta^2}\pm\sqrt{r_b^2-x_2^2-(x_1-r_c\cos \theta)^2}.
\end{equation}
$I_{q}$ is the regularized Beta function, $x_1$ and $x_2$ represent the projected offset from the centre of the cluster in the $x$ and $y$ directions, and 
\begin{equation}
q_i(z)=z^2/(r_{c,i}^2+x_1^2+x_2^2+z^2).
\end{equation}
Following the formalism described above we simulate a cluster based loosely on the well studied Perseus cluster. For the large scale cluster electron gas density distribution we use a two component $\beta$ model, the details of which are tabulated in Table~\ref{tab:cluster}. To this gas distribution we add three spherical bubbles of varying sizes and distances from the centre of the cluster. These bubbles are parameterized as shown in Fig.~\ref{fig:bubgeom} with values listed in Table~\ref{tab:bubbles}. The resulting brightness temperature distribution is shown in Fig.~\ref{fig:bubbles}(a), as is a cut through the largest bubble in Fig.~\ref{fig:bubbles}(b).
\begin{table}
\caption{Parameters of the double $\beta$ model used to simulate the Perseus-like cluster (see Pfrommer et~al. 2005). \label{tab:cluster}}
\begin{center}
\smallskip
\begin{tabular}{lcccc}
\hline\hline
Component & $r_c$ & $n_0$ & $\beta$ & $T_{\rm{e}}$ \\
          & (kpc) & (cm$^{-3}$) & & (keV) \\
\hline
1 & 47 & 0.0043 & 0.94 & 2.56 \\
2 & 178 & 0.046 & 0.58 & 7.5 \\
\hline
\end{tabular}
\end{center}
\end{table}
\begin{table}
\caption{Positions and sizes of the three cluster bubbles/cavities. The geometry of a bubble is illustrated in Fig.~\ref{fig:bubgeom}\label{tab:bubbles}}
\begin{center}
\smallskip
\begin{tabular}{lcccc}
\hline\hline
Bubble & $r_c$ & $r_b$ & $\theta$ & $\phi$ \\ 
& (kpc) & (kpc) & (deg) & (deg)\\
\hline
1 & 12 & 8 & 50 & 90\\
2 & 18 & 5 & 150 & 30 \\
3 & 25 & 3 & $-50$ & 10\\
\hline
\end{tabular}
\end{center}
\end{table}

\section{Interferometric Simulations}
\label{sec:intsim}

The ALMA telescope is currently under construction on the Chajnantor plane in the Chilean Andes at an altitude of approximately 5000\,m. When finished it will comprise two arrays: the first of up to fifty four 12\,m dishes and the second a close packed array of twelve 7\,m antennas. Although it was originally designed to span the frequency range 31--950\,GHz the telescope was rescoped in 2001 and a decision was made to delay the two lowest frequency bands (31--84\,GHz) beyond the start of science operations. Recently there has been renewed international interest in Band 1 (31--45\,GHz) being made a high priority (Johnstone et~al. 2009). 

\begin{figure}
\centerline{\includegraphics[width=0.53\textwidth]{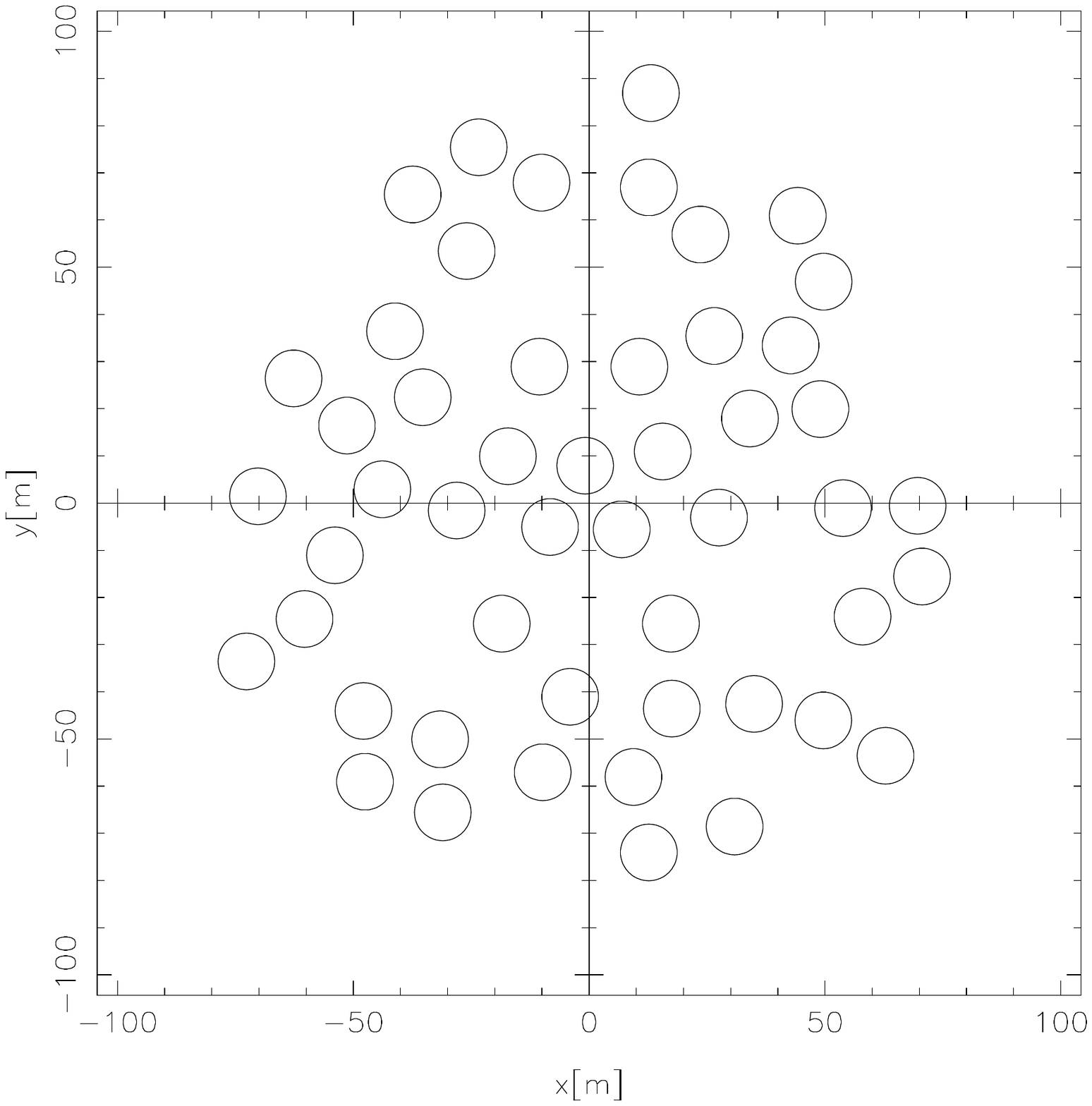}\qquad\includegraphics[width=0.5\textwidth]{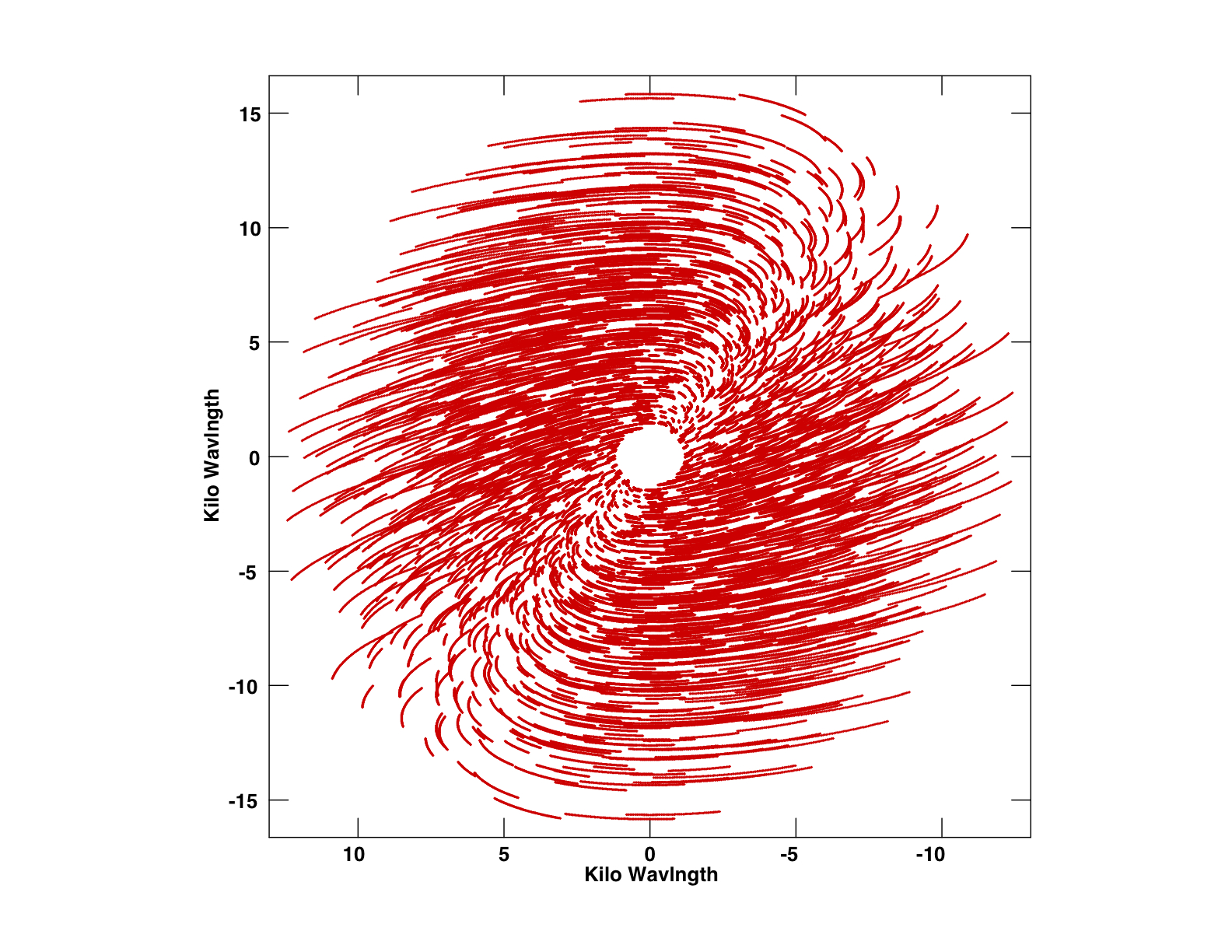}}

\centerline{(a) \hspace{0.5\textwidth} (b)}
\caption{ (a) ALMA telescope geometry and (b) \emph{uv} coverage of ALMA towards a source at zenith (RA$=12^{\rm{h}},\,\,\delta=-23^{\circ}$).\label{fig:geom}}
\end{figure}

There are a number of proposed geometries for the 12\,m dishes. In this work we utilise the most compact proposed distribution, see Fig.~\ref{fig:geom}(a), which has fifty dishes. We use this geometry to simulate an observation of a Perseus-like cluster at the zenith of the telescope ($\delta=-23^{\circ}$) and the \emph{uv} coverage for this observation is show in Fig.~\ref{fig:geom}(b). The synthesized beam from this \emph{uv} coverage is $14''\times10''$. 

The interferometric simulations described here were performed using the custom software package {\sc profile} which Fourier transforms an input map (in this case our simulated Perseus-like cluster) and then samples it at the \emph{uv} positions derived from the telescope geometry and a Right Ascension and Declination. The sampling is performed at specified time increments over a user defined range of Hour Angle. The sampled visibilities are then outputted in the standard \emph{uv} {\sc fits} format. The {\sc fits} file is then imaged and deconvolved using standard routines in the widely used {\sc aips} data analysis software package. In what follows the data has been mapped and {\sc clean}ed using the routine {\sc imagr}.

\begin{figure}
\centerline{ \includegraphics[width=0.5\textwidth]{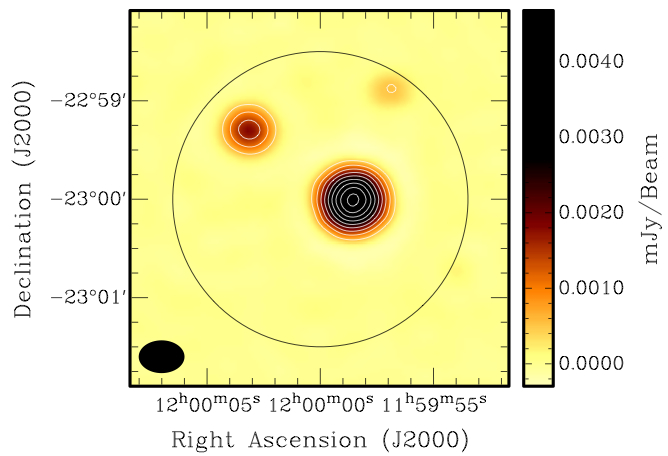}\qquad \includegraphics[width=0.5\textwidth,height=0.34\textwidth]{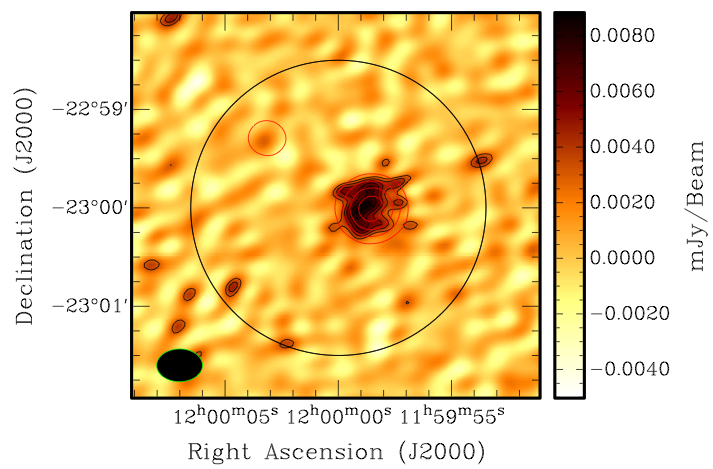} }
\centerline{ (a) \hspace{0.5\textwidth} (b) }
\caption{Simulated cluster. (a) shows the model cluster in Fig.~\ref{fig:bubbles} as it would be seen by the ALMA telescope in its most compact configuration. 	This simulated observation is noiseless and has no primary beam applied. (b) shows the model cluster as it would be observed in 32\,hr with the ALMA telescope. This simulated observation has noise of 7\,mJy/$\sqrt{\rm{s}}$. The solid circle in both images shows the FWHM of the ALMA Band 1 primary beam ($\approx3'$).\label{fig:nn}}
\end{figure}

We simulate an observation of our Perseus-like cluster in two ways. Firstly we simulate the data without applying a primary beam taper to the map and without adding noise to the simulated visibilities. This type of simulation allows us to examine which features are recovered by the \emph{uv} coverage of ALMA without corrupting them. This simulation is shown in Fig.~\ref{fig:nn} where we can immediately see that the large scale structure of the cluster has not been recovered by the baseline distribution of the fifty ALMA dishes. This is expected since the \emph{uv} coverage of ALMA lacks the short baselines necessary to recover this flux density. The small scale structure of the three bubbles appears clearly as positive features in the map, although the greyscale shows us that the surface brightness difference between the bubbles and the background is on the scale of $\mu$Jy. This allows us to estimate the integration time necessary to detect these features when the thermal noise of the telescope is considered. The thermal noise of ALMA is currently predicted to be 7\,mJy/$\sqrt{\rm{s}}$, and so at least 32 hours of observation will be required to detect the strongest of the features in this map.

The second method of simulation that we employ simulates more fully a true observation. In this instance we apply a Gaussian primary beam to approximate the aperture illumination function of the ALMA telescope. For the 12\,m dishes operating at 31\,GHz this primary beam will be $\approx 3'$ FWHM. Noise is added directly to the visibilities before gridding and imaging. Fig.~\ref{fig:nn}(b) shows one realization of this type of simulation, where noise appropriate to 32 hours of observation has been added to the visibilities. We note that these simulations include only the thermal cluster gas. We assume that any contaminating radio point sources within the field will have already been subtracted from the visibilities using the long baseline response. In addition we assume that any halo emission from a possible radio relic will be negligible at 30\,GHz.

\section{Discussion}
\label{sec:disc}

The resulting surface brightness map from these simulations differs considerably from that of Pfrommer et~al. (2005). The synthetic ALMA observations in Pfrommer et~al. were created by simply convolving the input cluster map with a Gaussian {\sc clean} beam, rather than simulating the observing response of an interferometer, and therefore retaining the large scale structure. Our simulations show that considerably longer integration times will be required in order to recover the small scale structure within the cluster gas.

In addition, we note that these simulations are performed at 31\,GHz in order to make predictions for the proposed Band 1 ALMA receiver, rather than 144\,GHz as simulated by Pfrommer et~al. (2005). At this lower frequency the brightness temperature of the SZ effect in Kelvins is in fact three times brighter than at 144\,GHz. In terms of brightness temperature alone (assuming comparable bandwidths and system temperatures at the two frequencies), one could consider that this will then increase the required integration time calculated here by an additional factor of nine to detect the same structures at 144\,GHz. Against this, we will not have perfectly scaled arrays at the two frequencies and so we must also recognise that the surface brightness at 144\,GHz in Jy\,sr$^{-1}$ will in fact be a factor seven higher than at 31\,GHz. However one must also take into account the difference in the length of the baselines as measured in units of wavelength between the two frequencies. At 144\,GHz the baselines will be almost a factor of five longer in $\lambda$ and consequently less sensitive to the extended emission of the larger features; this decrease in sensitivity is generally taken to be $\propto \lambda^2$, and will therefore cancel out the relative increase in surface brightness. In short, for these extended features an increase in integration time of a factor of approximately nine is required at 144\,GHz relative to 31\,GHz.

Including data from the ALMA compact array will be beneficial in recovering additional signal from the more extended features in the cluster sub-structure as it will possess shorter baselines and hence more sensitivity to larger angular scales; however we note that the reduced collecting area of the 7\,m dishes will limit the sensitivity to weak signals. Observations using the larger collecting area of the 12\,m dish array are limited by their minimum baseline spacing. The geometry considered here for this array is already extremely compact for dishes of this dimension and consequently there is unlikely to be any further improvement in considering revised antenna arrangements. Such arrays with longer baselines are more suited to observing smaller features in the intra-cluster medium, however these features, unless elongated along the line of sight, are inherently fainter due to their shorter path-length and consequently the smaller integrated pressure difference they create.

\begin{figure}
\centerline{ \includegraphics[width=0.5\textwidth]{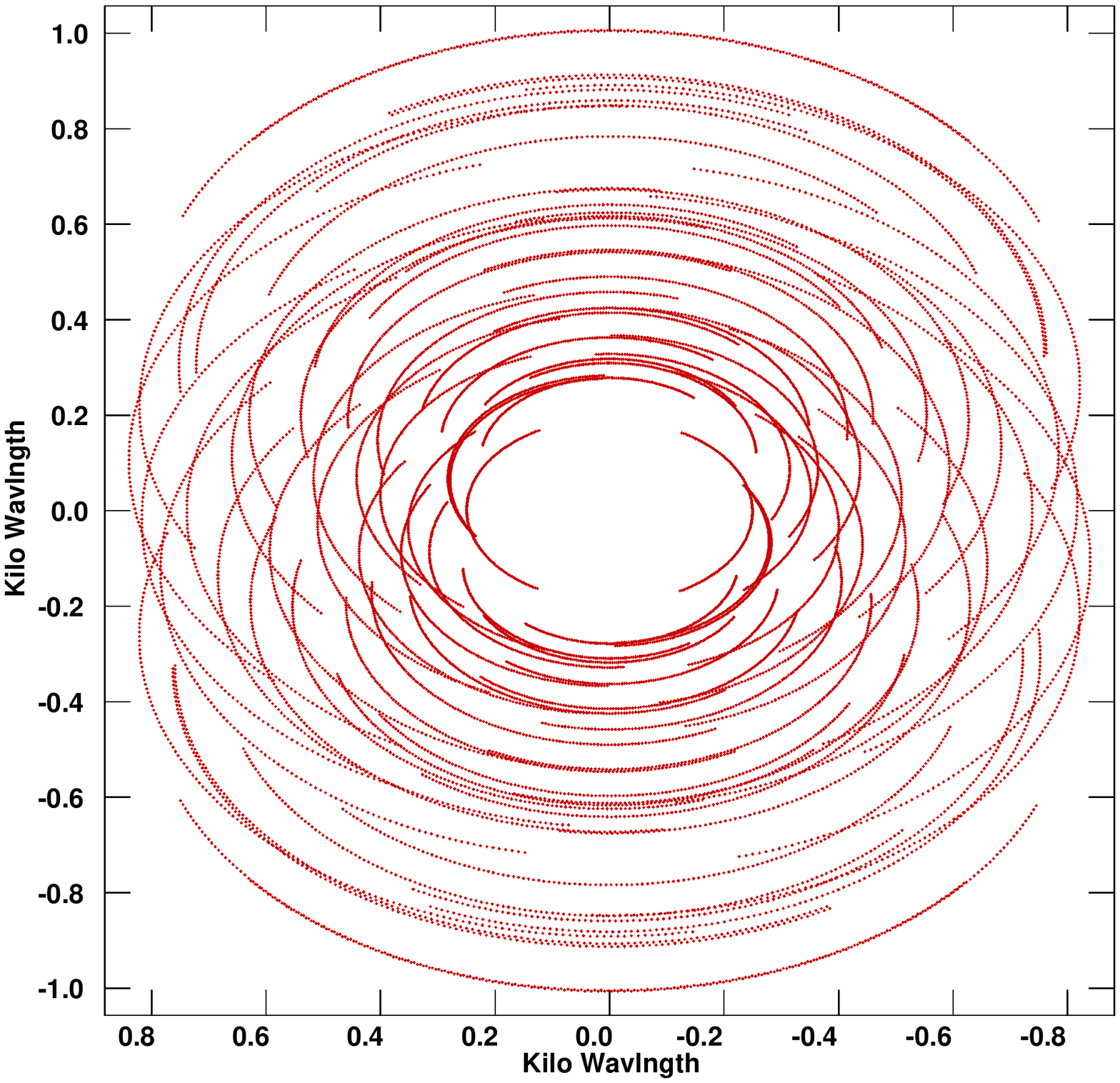}\qquad \includegraphics[width=0.5\textwidth]{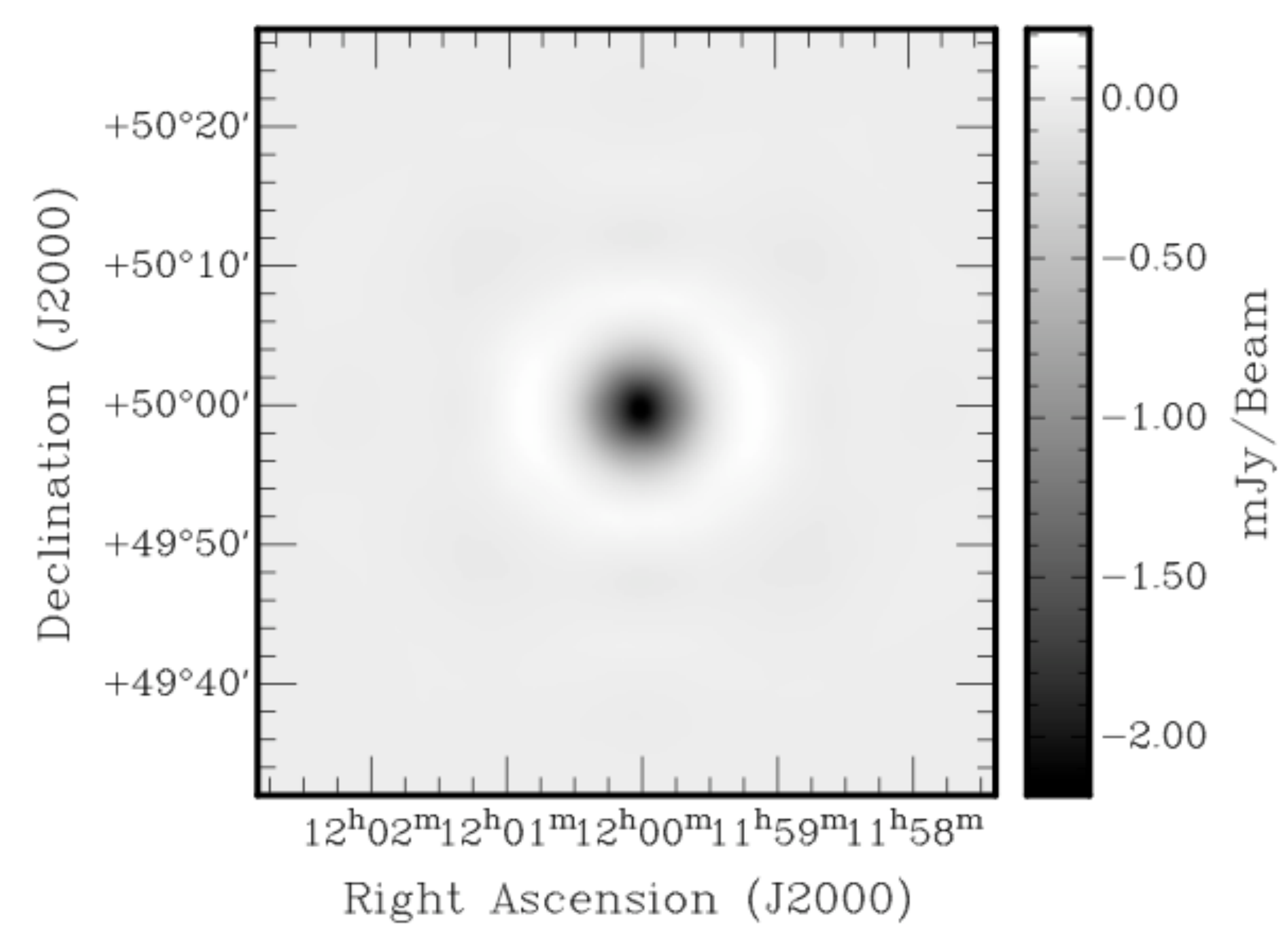}}
\caption{(a) \emph{uv} coverage for the AMI telescope, see text for details and (b) simulated observation of the Perseus-like cluster for the AMI telescope. \label{fig:ami}}
\end{figure}

For comparison with current SZ instruments we present simulated data for the Perseus-like cluster observed with the Arcminute Microkelvin
Imager (AMI) telescope in Cambridge, UK (AMI Consortium: Zwart et~al. 2008). The AMI telescope has been designed specifically to detect
clusters of galaxies through the SZ effect and is currently completing a blind survey for such clusters (Kneissl et~al. 2001). The \emph{uv}
coverage shown in Fig.~\ref{fig:ami} covers a range from approximately $200-1000\lambda$, containing a far greater number of short spacing
than the ALMA coverage shown in Fig.~\ref{fig:geom}, but lacking the coverage at spacings greater than $1000\lambda$. This implies, and is
demonstrated in Fig.~\ref{fig:ami}(b), that the AMI telescope will be sensitive only to the larger scale structure of the Perseus-like cluster. Since the majority of the flux density for such a cluster is contained in the large scale structure the surface brightness and recovered flux density for this observation is much higher than that of the ALMA observation.

\section{Conclusions}

In conclusion, we have shown that a Band 1 receiver system for the ALMA telescope will be capable of providing observational data which allows examination of the small scale structure within the intra-cluster gas of clusters of galaxies. Information on these scales is not available from present SZ telescopes which are optimized to recover the larger scale structure of such clusters. We have demonstrated that the surface brightness levels which will be recovered from these observations are smaller than previously calculated and to detect cavities such as those seen within the Perseus cluster will require at least 32 hours of pointed observation with an ALMA array of fifty dishes.

\label{lastpage}
\end{document}